\documentclass[%
 amsmath,amssymb,
 aps,
 pra,
twocolumn,
 floats,
]{revtex4-1}
\usepackage{epsfig}
\usepackage{graphicx}
\usepackage{dcolumn}
\usepackage{bm}

\begin{document}

\preprint{Anomalous AC Stark}
\title{Anomalous light shift through quantum jumps in quasi-resonant\\ Rayleigh scattering}
\author{D. G. Norris}
\author{A. D. Cimmarusti}
\author{L. A. Orozco}
 \email{lorozco@umd.edu}
\author{P. Barberis-Blostein}%
 \altaffiliation[Also at ]{Instituto de Investigaci{\'o}n en Matem{\'a}ticas Aplicadas y en Sistemas, Universidad Nacional Aut{\'o}noma de M{\'e}xico, M{\'e}xico, DF 01000, M{\'e}xico.}
\author{H. J. Carmichael}
 \altaffiliation[Also at ]{Department of Physics, University of Auckland, Private Bag 92019, Auckland, New Zealand.}
\affiliation{Joint Quantum Institute, Department of Physics, University
of Maryland and National Institute of Standards and Technology, College Park, MD 20742-4111, U.S.A.}%

\date{\today}

\begin{abstract}
An anomalous light shift in the precession of a ground-state Zeeman coherence is observed: the Larmor frequency \emph{increases} with the strength of a drive that is blue (red) detuned from a transition out of the lower (upper) energy level.  Our measurements are made on $^{85}{\rm Rb}$ atoms traversing an optical cavity containing a few photons; shifts as large as 1\% per photon are recorded. The anomalous shift arises from an accumulation of phase driven by quantum jumps. It is stochastic and accompanied by broadening.
\end{abstract}

\maketitle

Elastic Rayleigh scattering \cite{bermanbook11} is a ubiquitous process in the manipulation of atoms by light, and a widely used tool for the projective measurement of their quantum states. While its basic properties have long been known, subtleties continue to be uncovered. Uys {\it et al.}~\cite{uys10} have recently shown that its contribution to the decoherence rate of a ground-state superposition (two-level system) is given by the square of the difference of the scattering amplitudes; if the amplitudes interfere constructively, there is decoherence even when the rates of scattering from the two levels of the superposition are equal, contrasting previous work \cite{ozeri05} that found such effects negligible.

The elastic Rayleigh scattering considered by Uys \emph{et al.} is exemplified by measurements on a $^9{\rm Be}^+$ ion, with a ground-to-excited-state detuning of tens of GHz, far in excess of the excited state linewidth. In this Letter, we report on an anomalous light shift observed in a quasi-resonant system of ${\rm ^{85}Rb}$ atoms. This shift, which reverses the sign of the usual AC Stark shift, is similarly due to elastic Rayleigh scattering and driven by precisely the same mechanism as the decoherence reported in \cite{uys10}. Most generally, decoherence and shift exist side by side, with the decoherence dominant far from resonance and the anomalous light shift dominant in the quasi-resonant regime. The two sides are unified by an analysis of the quantum-jump-driven evolution of coherence under elastic Rayleigh scattering.

Our experimental observation is made through conditional detection of photons scattered from ${\rm ^{85}Rb}$ atoms into an optical cavity mode at moderate-to-weak dipole coupling strengths \cite{norris10}. Detection of a first photon creates coherence between Zeeman-shifted ground states, $|g_-\rangle$ and $|g_+\rangle$, where the Zeeman splitting is significantly smaller than the excited state linewidth and the cavity width. Evolution of this coherence is observed as a quantum beat written on the probability of a subsequent second photon detection \cite{schubert95}. The beat note then reveals the anomalous shift: the observed Larmor frequency \emph{increases} with the strength of a drive that is blue (red) detuned from the optical transition out of the lower (higher) ground state, opposite to what one would naively predict from the AC Stark effect applied to each ground-state level. We first explain the origin of this anomalous shift and its connection to the decoherence reported in \cite{uys10} using a simplified model. We then present our experimental measurements, which we compare with full quantum trajectory simulations. A complementary master equation treatment will appear elsewhere.

\begin{figure}
\includegraphics[width=0.9\linewidth]{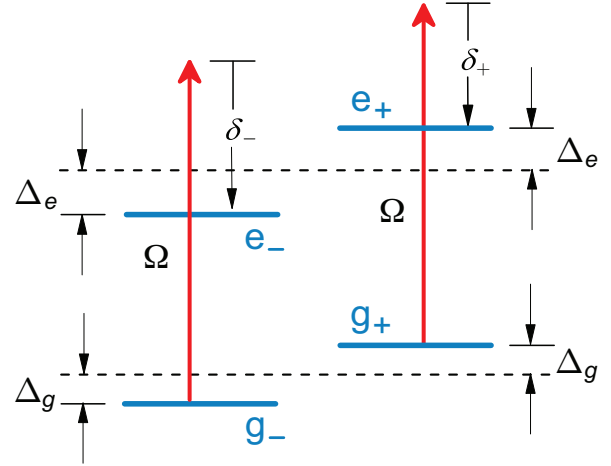}
\caption{\label{atom-structure} (color online) Four-level atom with unequal Zeeman shifts in the ground ($\Delta_g$) and excited ($\Delta_e$) states. A coherent field $\Omega$ is detuned by $\delta_0=(\delta_-+\delta_+)/2$ from the splitting between the unshifted levels (dashed lines), driving both transitions simultaneously. All detunings are defined as a transition frequency minus the laser frequency. Zeeman shifts are positive by definition.}
\label{model}
\end{figure}

Using quantum trajectories \cite{carmichael93book,plenio98}, we follow the evolution of a ground-state coherence, $|\psi_g\rangle =(|g_-\rangle+|g_+\rangle)/\sqrt2$, initially created between $m=\pm1$ Zeeman sublevels. Neglecting $\sigma$ spontaneous emissions allows us to treat the closed four-level model of Fig.~\ref{model}. (In practice the four-level atom of Fig.~\ref{model} is an approximation to a sub-manifold that shuttles backwards and forwards within the large angular momentum manifold of our experimental system as we track the coherence through many optical pumping events. The approximation neglects changes in detuning and Clebsch-Gordan coefficients for some scattering cycles.  It is nevertheless a good one because we are considering time scales short compared with the long-time optical pumping limit.)  A weak $\pi$-polarized coherent drive, Rabi frequency $2\Omega$, has detunings $\delta_\pm=\delta_0\pm(\Delta_e-\Delta_g)$ from the $|g_\pm\rangle\to|e_\pm\rangle$ Zeeman-shifted transitions; $\delta_0$ is the detuning from the unshifted transition and $\Delta_g$ ($\Delta_e$) are unequal ground-state (excited-state) Zeeman shifts. The evolution between Rayleigh scattering events (quantum jumps) follows equations of motion for Schr\"odinger amplitudes:
\begin{eqnarray}
\dot{c}_{e\pm}&=&-\frac{\gamma}{2}c_{e\pm}\mp i\Delta_{e}c_{e\pm}+\Omega e^{i\delta_0t}c_{g\pm},\nonumber\\
\dot{c}_{g\pm}&=&\mp i\Delta_{g}c_{g\pm}-\Omega e^{-i\delta_0t}c_{e\pm},
\label{between_jump_eqs}
\end{eqnarray}
written in a frame rotating at the frequency of the atom, where $\gamma$ is the excited state decay rate. After solving this system to lowest order in $(2\Omega/\gamma)^2$ we arrive at the evolving ground-state superposition,
\begin{equation}
|\psi_g(t)\rangle=\frac1{\sqrt2}[\alpha_-(t)|g_-\rangle+\alpha_+(t)|g_+\rangle],
\label{initial_ground-state_coherence}
\end{equation}
with amplitudes
\begin{equation}
\alpha_\pm(t)=\exp(\mp i\Delta_gt)\exp(-i\delta_{AC}^{(\pm)}t)\exp(-\gamma^{(\pm)}t),
\label{between_jump_amplitudes}
\end{equation}
which, through the first of Eqs.~(\ref{between_jump_eqs}), drives a steady-state superposition in the excited state ($t\gg\gamma^{-1}$):
\begin{equation}
\label{excited-state_coherence}
|\psi_e(t)\rangle = e^{i\delta_0t}\frac{\Omega}{\sqrt{2}}[A^{(-)}\alpha_-(t)|e_-\rangle+A^{(+)}\alpha_+(t)|e_+\rangle],
\end{equation}
where we have the definitions
\begin{equation}
\delta_{AC}^{(\pm)}=-\frac{\delta_{\pm}}{\gamma/2}\gamma^{(\pm)},\qquad\gamma^{(\pm)}=\frac12\Omega^2\gamma|A^{(\pm)}|^2,
\label{AC_Stark_shifts&rates}
\end{equation}
with scattering amplitudes
\begin{equation}
A^{(\pm)}=\frac1{\gamma/2+i\delta_{\pm}}.
\label{scattering_amplitudes}
\end{equation}
Evolution of the amplitudes $\alpha_\pm(t)$ is composed of three pieces: (i) phase accumulation, $\pm\Delta_g t$, from the unshifted Larmor precession, (ii)  phase accumulation, $\delta^{(\pm)}_{AC}t$, from the AC Stark shifts, and (iii) damping at rates $\gamma^{(\pm)}$, which accounts for backaction from the null measurement when the scattering rates on the two components of the superposition are unequal---the component with smaller (larger) scattering rate grows (shrinks) relative to the other. Note that for resonant driving of the unshifted transition ($\delta_0=0$) and $\Delta_e>\Delta_g$, the AC Stark shifts
\begin{equation}
\delta^{(\pm)}_{AC}=\mp(\Delta_e-\Delta_g)\frac{\Omega^2}{(\gamma/2)^2+(\Delta_e-\Delta_g)^2}
\label{AC_Stark shifts}
\end{equation}
\emph{reduce} the Larmor precession frequency, contrary to what we observe.

Now consider the effect of a quantum jump (a Rayleigh scattering event on the driven transition), which transfers the excited state amplitudes [Eq.~(\ref{excited-state_coherence})] to the ground state; Eq.~(\ref{initial_ground-state_coherence}) is replaced by
\begin{equation}
|\psi_g(t)\rangle=\frac{\Omega}{\sqrt2}[A^{(-)}\alpha_-(t)|g_-\rangle+A^{(+)}\alpha_+(t)|g_+\rangle],
\label{new_ground-state_coherence}
\end{equation}
with the insertion of the scattering amplitudes $A^{(\pm)}$. The evolution of the ground-state coherence acquires two new features: (iv) an additional phase advance, ${\rm arg}(A^{(\pm)})$, and (v) an additional change of relative amplitudes due to backaction---the component of the superposition with smaller (larger) scattering rate shrinks (grows) relative to the other. As the Rayleigh scattering proceeds, one scattered photon after another, this evolution through quantum jumps is compounded; after $n$ scattering events, Eq.~(\ref{new_ground-state_coherence}) carries over with $\Omega\to\Omega^n$ and $A^{(\pm)}\to(A^{(\pm)})^n$.

When the rates of scattering on the two components of the superposition are unequal, quantum trajectories effect a measurement of the spin state. Processes (iii) and (v) compete, such that a fluctuation in the number of scattered photons makes the amplitude of one component grow relative to the other---an excess (deficit) of photons scattered brings localization on the spin state with larger (smaller) scattering rate. This corresponds to the usual case of projective state measurement through Rayleigh scattering.

\begin{figure}[b]
\includegraphics[width=0.9\linewidth]{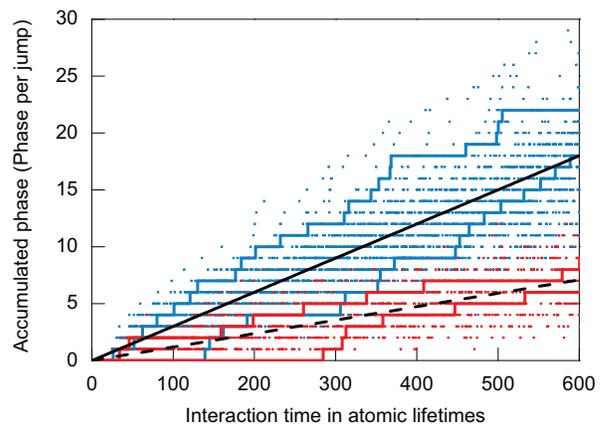}\caption{\label{qmc}  (color online) Sample quantum trajectories showing the phase accumulated in the ground state through process (iv) of the main text for $\Omega=0.075\gamma$ (red) and $0.125\gamma$ (blue), with $\delta_+=-\delta_-=0.5\gamma$. Black lines show the mean phase accumulation. Dots show the scatter over 200 runs.}
\label{sample_trajectories}
\end{figure}

The evolution is quite different when the scattering rates are equal ($\delta_-=-\delta_+=\Delta_e-\Delta_g$), as phase accumulation through process (iv) is  now  important. Uys \emph{et al.}~\cite{uys10} examined this case for far-off-resonance driving ($\delta_-=-\delta_+\gg\gamma/2$). In this regime, $A^{(-)}/A^{(+)}=-1$, and each quantum jump shifts the phase of the superposition by $\pi$, leading to their reported decoherence. In contrast, here we consider a drive close to resonance ($\delta_-=-\delta_+\ll\gamma/2$), where the phase shift per jump is small, and a significant change of phase accumulates only slowly after many photons are scattered. The evolution approaches the phase diffusion process of Fig.~\ref{sample_trajectories}; there is a mean accumulation of phase (black lines) and a spread leading to decoherence. So long as the spread is smaller than $\pi$, the mean accumulation is an effective frequency shift.

For concrete results we sum over trajectories. At weak drive, this amounts to a sum over a Poisson distribution of photons scattered in time $t$. We find diagonal density matrix elements
\begin{equation}
\rho_{g_\pm,g_\pm}=\frac12e^{-2\gamma^{(\pm)}t}\sum_{n=0}^\infty\frac{[\gamma\Omega^2|A^{(\pm)}|^2t]^n}{n!}=\frac12,
\label{diagonal_matrix_elecments}
\end{equation}
and, for the off-diagonal coherence,
\begin{eqnarray}
\rho_{g_-,g_+}=\frac12e^{-(\Gamma/2)t}e^{i2(\Delta_g+\Delta)t},
\label{off-diagonal_coherence}
\end{eqnarray}
with decoherence rate
\begin{equation}
\Gamma/2=\gamma^{(+)}+\gamma^{(-)}-\Omega^2\gamma{\rm Re}[A^{(-)}(A^{(+)})^*]
\label{decay_rate}
\end{equation}
and net frequency shift
\begin{equation}
2\Delta=\delta_{AC}^{(+)}-\delta_{AC}^{(-)}+\Omega^2\gamma{\rm Im}[A^{(-)}(A^{(+)})^*].
\label{frequency_shift}
\end{equation}
From Eqs.~(\ref{AC_Stark_shifts&rates}) and (\ref{decay_rate}), $\Gamma=\Omega^2\gamma|A^{(+)}-A^{(-)}|^2$, which agrees with the decoherence rate associated with elastic scattering reported in \cite{uys10} [Eq.~(7)]. Near resonance $\Gamma=64\Omega^2|\delta_\pm|^2/\gamma^3$, and the decoherence is sufficiently small to speak of a well-defined frequency shift. The total shift comprises two terms: an AC Stark shift, $\delta_{AC}^{(+)}-\delta_{AC}^{(-)}$, and an additional contribution due to quantum jumps, $\Omega^2\gamma{\rm Im}[A^{(-)}(A^{(+)})^*]=16\Omega^2|\delta_\pm|/\gamma^2$. The shift due to jumps has twice the magnitude of the AC Stark shift and opposite sign; the usual light shift is effectively reversed. We refer to an ``anomalous'' light shift in analogy with anomalous dispersion, which also occurs near resonance.

We draw an important distinction between this phenomenon, which arises from the repeated collapse of a ground-to-excited state coherence, and the similar effect caused by essentially classical frequency switching \cite{dykman10}. In the early 1960s, prior to the use of lasers, Cohen-Tannoudji and Kastler  studied the effects of an \emph{incoherent} optical drive on ground-state coherences in $^{199}$Hg atoms using nuclear magnetic resonance techniques \cite{cohen62,kastler63}. They observed small ($\sim1\mkern2mu{\rm Hz}$) power-dependent shifts and broadenings of the precession frequency, classifying the shifts according to two kinds: \emph{shifts due to virtual transitions} (AC Stark shifts) and \emph{shifts due to real transitions}. The latter arise when an atom is excited and precesses for some time with the excited-state Larmor frequency before returning to the ground state. These are also, in effect, shifts due to quantum jumps, though different in character and degree from those presented in this Letter. With incoherent drive, the Larmor frequency changes abruptly from 2$\Delta_g$ to 2$\Delta_e$ for a lifetime $1/\gamma$, giving a phase advance per up-down cycle of $2(\Delta_e-\Delta_g)/\gamma$.  This is smaller by a factor of two than the case of coherent excitation, where the phase advance per down-jump (there is no up-jump) is $-i{\rm ln}(A^{(+)}/A^{(-)})=4(\Delta_e-\Delta_g)/\gamma$. Remarkably, the net shift in the two cases agrees exactly at resonance ($\delta_0=0$) due to the contribution from the AC Stark shift for coherent drive (there is no AC Stark shift for broadband excitation near resonance). More generally, however, the shifts are different when the drive is detuned.

\begin{figure}
\includegraphics[width=0.9\linewidth]{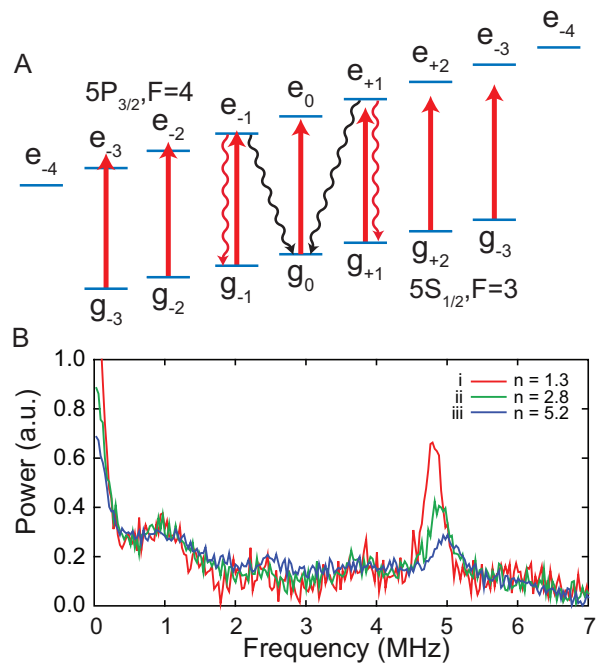}\caption{ (color online) A:  Level structure of $^{85}$Rb in a weak magnetic field. The drive is $\pi$ polarized while spontaneous emission returns atoms to the ground state through either $\pi$ or $\sigma$ transitions. B: FFT of the ground-state beat for three different drive strengths (intracavity photon numbers as noted).}
\label{levels-fft}
\end{figure}

Turning now to our experiment, we follow the evolution of a spontaneously generated coherence \cite{javanainen92} stored in the ground-state of $^{85}$Rb atoms. The atoms are prepared in the $g_0$ magnetic sublevel (Fig.~\ref{levels-fft}A) and traverse an optical cavity as a dilute atomic beam with mean speed $\sim20\mkern2mu{\rm ms}^{-1}$. Inside the cavity, they couple to vertically and horizontally polarized cavity modes at moderate-to-weak dipole coupling strength. The vertical polarization is oriented parallel to a weak magnetic field and the cavity is weakly driven with vertically polarized coherent light, thus driving $\pi$ transitions in the atoms (red arrows in Fig.~\ref{levels-fft}A). The cavity and drive are resonant with the $g_0\to e_0$ atomic transition.

A ground-state coherence, $|\psi_g\rangle=(|g_{-1}\rangle +|g_{+1}\rangle)/\sqrt2$, is spontaneously created each time an atom scatters a photon via its $e_0$ excited state into the horizontally polarized cavity mode. The scattered photon is distinguished from drive photons by means of polarization filtering at the cavity output, and its detection initiates subsequent monitoring of the time evolution of the coherence created. This monitoring is enabled by a quantum beat, which is written on the probability for the scattering of a second horizontally polarized photon (the origin is the interference of emission pathways shown by the black wavy lines in Fig.~\ref{levels-fft}A); thus, the phase accumulation of the coherence appears as the phase accumulation of a quantum beat, recorded in the intensity correlation function of horizontally polarized photons scattered via the cavity mode. The cavity provides excellent mode quality for this monitoring of the dynamics. Further background on our experimental method appears in Ref.~\cite{norris10}.

The frequency of the quantum beat directly measures the oscillation of the coherence in Eq.~(\ref{off-diagonal_coherence}), and hence the net frequency shift in Eq.~(\ref{frequency_shift}). The shift due to quantum jumps is caused by the additional scattering of photons, through spontaneous emission, to modes \emph{other than the cavity modes}. Figure~\ref{levels-fft}B shows three examples of the Fast Fourier Transform (FFT) of the measured quantum beat. From these spectra, we extract a center frequency and full-width-at-half-maximum (FWHM). Clearly, as the drive strength increases, the center frequency shifts up, not down as from the effect of the AC Stark shift, and the spectrum broadens. These results are in qualitative agreement with the theoretical prediction from our simplified model.

\begin{figure}
\includegraphics[width=0.8\linewidth]{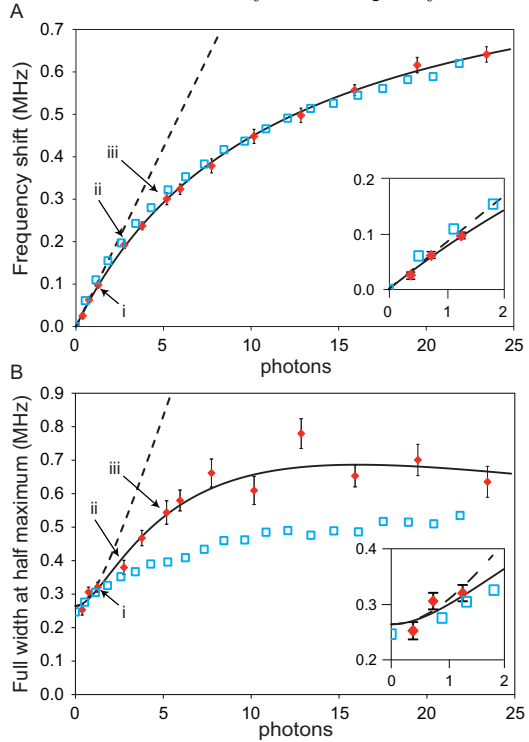}\caption{(color online) Quantum beat frequency and width as a function of drive photon number in a $5\mkern2mu{\rm G}$ magnetic field. A: Shift of the FFT center frequency and B: FFT FWHM, from measurements (red diamonds) and simulations (blue squares). Solid (Dashed) lines are least-squares fits to the data with (without) atomic saturation. Labeled points correspond to Fig~\ref{levels-fft}B. Insets show low intensity behavior.
\label{analysis}}
\end{figure}

For a more careful comparison with theory we must account for the main experimental complications.  These include optical pumping within the full level structure of $^{85}$Rb (Fig.~\ref{levels-fft}A), the finite interaction time and unequal coupling strengths associated with an atomic beam, and saturation.  To this end we expand the previous quantum trajectory treatment as a full quantum Monte Carlo simulation (see Ref.~\cite{norris10} for details).

Figure \ref{analysis}A shows the measured shift of the FFT center frequency from the natural Larmor frequency ($4.7\mkern2mu{\rm MHz}$ in a $5\mkern2mu{\rm G}$ magnetic field) as a function of photon number, $n=(\Omega/g)^2$, $g$ the dipole coupling strength, in the driven cavity mode. Our theoretical simulation achieves excellent quantitative agreement with the data.  Despite complications brought by the level structure and unequal coupling strengths, the high intensity regime is well-described by allowing for saturation of the jump rate as in a two-level atom, i.e., $4\Omega^2/\gamma\to(4\Omega^2/\gamma)(1+8\Omega^2/\gamma^2)^{-1}$; the solid line is a least-squares fit to this functional form. Low drive gives a linear slope of $100\mkern2mu{\rm kHz}$ per intracavity photon.

Figure \ref{analysis}B shows the FFT FWHM as a function of the number of drive photons in the cavity.  The intercept at zero photons ($0.27\mkern2mu{\rm MHz}$) arises from the transit time of atoms through the cavity mode. The increased width at higher drive is a result of the jump-induced phase diffusion illustrated in Fig.~\ref{sample_trajectories}.  The solid line is a fit to the data, combining the two linewidths with saturation in the jump rate. While the trend is the same---increasing FWHM with increasing number of drive photons, plus saturation---there is clearly a discrepancy between the measurements
and the simulation. We expect there is an unaccounted for noise in the experiment that increases the rate of decoherence. A likely candidate is imperfect preparation of the $g_0$ ground state.

We have studied the effects of Rayleigh scattering on a ground-state superposition in the regime of equal scattering rates. We showed that the decoherence reported for a far-from-resonance drive in \cite{uys10} evolves into an anomalous light shift (shift of opposite sign) when the scattering is quasi-resonant. The two effects are unified through their common mechanism, whereby phase is accumulated as quantum jumps map the ground state coherence when a photon is scattered. We have experimentally observed anomalous shifts that are five orders of magnitude larger than related ``jump shifts'' observed with thermal light \cite{cohen62,kastler63,bulos71}. Such shifts may need to be considered, for example, in the spectroscopy of multilevel atoms where interference is involved \cite{sansonetti11}. Our results further suggest that, given appropriate design, the dissipative process of spontaneous emission might itself be harnessed as a tool for coherent control \cite{wisemanbook}.

Work supported by NSF, CONACYT, M{\'e}xico, and the Marsden Fund of RSNZ. We thank J. J. Bollinger for useful discussions.

\end{document}